\documentclass[10pt,paperpaper]{article}
\usepackage{opex3}
\usepackage{cite}
\usepackage{bm}
\usepackage{mathrsfs}
\usepackage{amssymb,amsmath}

\usepackage{graphicx}
\usepackage{epsfig}
\usepackage{verbatim}

\usepackage{amsfonts}

\providecommand{\U}[1]{\protect\rule{.1in}{.1in}}

\begin{document}

\title{Classical to quantum transfer of optical vortices}

\author{Ver\'{o}nica Vicu\~na-Hern\'{a}ndez,$^{1}$ H\'ector Cruz-Ram\'irez,$^{1}$ Roberto Ram\'irez-Alarc\'on,$^{1}$ and Alfred B. U'Ren$^{1*}$}

\address{$^1$Instituto de Ciencias Nucleares, Universidad Nacional
Aut\'onoma de M\'exico, apdo. postal 70-543, 04510 D.F., M\'exico\\
}

\email{$^*$alfred.uren@nucleares.unam.mx} 



\begin{abstract}
We show that an optical vortex beam, implemented classically, can be transferred to the transverse amplitude of a heralded single photon.   For this purpose we have relied on the process of
spontaneous parametric downconversion (SPDC) for the generation of
signal and idler photon pairs, using a pump in the form of a Bessel-Gauss (BG) beam with orbital angular momentum (specifically, with topological charge $l=1$ and $l=2$).
We have designed our source so that it operates within the short SPDC crystal regime for which, the amplitude and phase of the pump may be transferred  to 
a heralded single photon.  In order to verify the vortex nature of our heralded single photon, we have shown that the conditional angular spectrum
and the transverse intensity at the single-photon level match similar measurements carried out for the pump.  In addition, we have shown 
that when our heralded single photon is diffracted through a triangular
aperture, the far-field single-photon transverse intensity exhibits the expected triangular arrangement
of intensity lobes associated with the presence of orbital angular momentum.
\end{abstract}



\section{Introduction}

The ability to prepare single photons with a well-defined transverse shape represents a crucial enabling technology on a number of fronts.     Large-alphabet quantum key distribution schemes based on the transverse shape of single photons~\cite{mirhosseini14} and the coherent transfer of orbital angular momentum (OAM) between light and matter for storage and retrieval of quantum information~\cite{barreiro03} constitute relevant examples.   Our key objective in this paper is to show that in the process of spontaneous parametric downconversion (SPDC), a complex transverse spatial structure -including phase structure- in the pump beam may be transferred to a signal-mode single photon, as heralded by the corresponding idler photon.  While the approach exploited here is general and could be applied to a pump field with an arbitrary transverse spatial structure, we have chosen for an experimental verification  a vortex pump in the form of a Bessel-Gauss (BG) beam carrying orbital angular momentum, specifically with topological charge $l=1$ and $l=2$. 

The physical principle on which our work is based is well known: in the SPDC process the angular spectrum of the pump can be transferred to one of the daughter photons, as heralded by the detection of the conjugate idler photon with a well defined transverse wavevector~\cite{monken98,koprulu11,walborn04}.   In~\cite{ramirez13} we have shown that this transfer can only take place for a sufficiently thin SPDC crystal, and/or for a pump beam which is sufficiently compact in transverse wavevector space.    It is interesting that within this short-crystal regime, the above transfer mechanism is in fact phase-preserving, so that it is the full \emph{transverse amplitude}, and not merely the transverse intensity (or angular spectrum) which is transferred.   This phase-preserving transverse amplitude transfer is indeed present implicitly in a number of works, e.g. in~\cite{koprulu11}.  In this paper we analyze some of the practical issues for the successful implementation of transverse amplitude transfer through heralding in SPDC.  We analyze on the one hand the effect of a non-vanishing angular acceptance of the heralding detector, and on the other hand the specific conditions under which the short-crystal approximation is valid as well as the effect of a departure from this regime.   Because an optical vortex is fundamentally defined by a phase $\exp(i l \phi)$, a vortex pump in the SPDC process is ideal to demonstrate transverse amplitude transfer to a heralded daughter photon.  In our experiment we verify that i) the single-photon angular spectrum, ii) the single-photon transverse intensity and  iii) the far-field diffraction pattern through a triangular aperture all match corresponding measurements carried out for the pump~\cite{hickmann10}.   Diffraction through a triangular aperture constitutes a novel method to ascertain the presence of OAM and for the determination of the topological charge, not used perviously to our knowledge for SPDC light.  In this paper we show for the first time to our knowledge that when using a vortex pump beam in the SPDC process, the transverse amplitude of the pump can be verifiably transferred to the signal-mode heralded single photon.  

Our motivation behind this work is that an essentially arbitrary  classical pump beam may be prepared relatively easily through the use of standard optical elements such as phase filters and spatial light modulators, which can then be transferred to a single photon through the approach demonstrated here.   There is a fundamental distinction with respect to other works where single photons are projected, \emph{post-generation}, to a state with a certain transverse spatial structure (e.g. in \cite{mair01}).  Indeed, a defining characteristic of our own work is that the photon pairs are generated already with the desired transverse structure, which is revealed in the signal-mode single photon transverse amplitude by the heralding process, and which in an ideal situation is an exact replica of that of the pump.

Within the realm of single-photon transverse spatial structure, the presence, manipulation and exploitation of OAM has generated a great deal of excitement~\cite{molinaterriza07}.
In the context of the SPDC process, in which individual pump photons are annihilated in a second-order nonlinear medium giving rise to the emission of signal and idler photon pairs, OAM has been used as a new degree of freedom in which to  demonstrate the existence of quantum entanglement~\cite{mair01}.    In this case, entanglement resides in the OAM / optical angle pair of conjugate variables, 
which can behave similarly to transverse momentum / position and frequency / time, in which Einstein Podolsky Rosen (EPR) correlations may occur~\cite{leach10}.    One of the most promising aspects of OAM photon-pair entanglement is that it opens up a clear road towards higher-dimensional entanglement since OAM represents a discrete but infinite-dimensional photonic degree of freedom~\cite{dada11}. 

Besides the presence of OAM, BG beams have some interesting properties which may be exploited at the single-photon level. On the one hand, they are more resistant, as compared to an equivalent Gaussian beam,  to deformation by transmission through a turbulent medium~\cite{tyler09,sheng13} or due to physical obstacles~\cite{mclaren14}.  On the other hand, they are non-diffractive over a certain propagation distance.  While in a previous work~\cite{cruz12} we have shown for a fundamental BG beam, i.e. without OAM, that this non-diffractive pump behavior can be ``inherited'' by the heralded single photons (a behaviour expected for BG beams of all orders), in this paper we have focused on pump to heralded single photon transverse amplitude transfer for higher-order BG beams which exhibit OAM.

\section{Theory}


We rely on the process of SPDC in a type-I, non-collinear, frequency-degenerate regime, in which individual pump photons are annihilated so as to  generate signal and idler photon pairs. The two-photon component of the state produced by SPDC, in the monochromatic pump limit (with pump frequency $\omega_p$), is given by~\cite{ramirez13}

\begin{equation}
|\Psi\rangle=\int d \textbf{k}^\bot_s \int d \omega_s \int d \textbf{k}^\bot_i A_s A_i S( \textbf{k}^\bot_s+ \textbf{k}^\bot_i) G( \textbf{k}^\bot_s,\textbf{k}^\bot_i,\omega_s)|\textbf{k}^\bot_s, \omega_s\rangle_s   |\textbf{k}^\bot_i, \omega_p-\omega_s\rangle_i \label{E:2phstate}
\end{equation}

\noindent where $|\textbf{k}^\bot, \omega \rangle_\mu\equiv a_\mu^\dagger(\textbf{k}^\bot, \omega) |0\rangle$ represents a single-photon Fock state characterized by  frequency $\omega$ and transverse wavector $\textbf{k}^\bot$,  with $\mu=s,i$ for the signal (s) and idler (i) photons respectively.   In Equation~\ref{E:2phstate}, $S( \textbf{k}^\bot)$ represents the pump transverse amplitude  (so that $|S( \textbf{k}^\bot)|^2$ is the pump angular spectrum) and $G( \textbf{k}^\bot_s,\textbf{k}^\bot_i,\omega_s)$ is the phase matching function given in terms of crystal properties including length, dispersion and walk off; its functional form, not needed for the current analysis, is given in our earlier paper \cite{ramirez13}.  $A_s$ and $A_i$ represent slow functions of the signal/idler frequencies~\cite{ramirez13}, respectively,  which may be approximated as constants for a limited detection bandwidth.

The heralding of a signal-mode, single photon by the detection of an idler photon with frequency $\tilde{\omega}_i$ and transverse wavevector $\tilde{\textbf{k}}^\bot_i$ may be represented by the action of the projection operator 

\begin{equation}
\hat{\Pi}( \tilde{\textbf{k}}^\bot_i )=| \tilde{\textbf{k}}^\bot_i ,\tilde{\omega}_i \rangle_i \langle  \tilde{\textbf{k}}^\bot_i ,\tilde{\omega}_i |_i
\end{equation}

\noindent  on the two-photon state $|\Psi \rangle$.   This results in a signal-mode, heralded single photon in a pure state given as


\begin{equation}
|\Psi(\tilde{\textbf{k}}^\bot_i) \rangle_s =\int d \textbf{k}^\bot_s S( \textbf{k}^\bot_s+  \tilde{\textbf{k}}^\bot_i) G( \textbf{k}^\bot_s,  \tilde{\textbf{k}}^\bot_i,\omega_p-\tilde{\omega}_i )| \textbf{k}^\bot_s,\omega_p-\tilde{\omega}_i \rangle_s. \label{E:transfer}
\end{equation}

Note that in a regime where the so-called short-crystal approximation is valid, i.e. for a sufficiently short crystal and/or for a sufficiently compact pump in transverse wavector space, the function $|G( \textbf{k}^\bot_s,\tilde{\textbf{k}}^\bot_i,\omega_p-\tilde{\omega}_i )|$ has a greater width than the function $|S( \textbf{k}^\bot_s+ \tilde{\textbf{k}}^\bot_i)|$, so that the former may be disregarded.  In this case, the transverse amplitude of the heralded single photon becomes identical to the pump transverse amplitude, except displaced by $-\tilde{\textbf{k}}^\bot_i$ in transverse wavevector space.    Importantly, in the heralding process the single-photon amplitude retains any phase structure present in the pump beam; this is evident in Eq.~(\ref{E:transfer}).   In the context of the subject of this paper, this means that a vortex structure $\mbox{exp}(i l \phi)$ in the pump will also appear in the single-photon amplitude.  We refer to this as transfer of the (classical) transverse amplitude of the pump to the transverse amplitude of the signal-mode single photon.    

Transverse amplitude transfer as discussed in the previous paragraph is based on the plane-wave projection of the idler photon on a single wavevector.    In the more general case where the angular acceptance of the idler detector is given by a function $g(\textbf{k}^\bot)$ and where the contributions from each idler wavevector within the acceptance function are summed incoherently (as is the case for our experimental conditions, see below), the two photon state becomes a statistical mixture of pure states of the type in Eq.~(\ref{E:transfer}), as follows

\begin{equation}
\hat{\rho}_s=\int d \tilde{\textbf{k'}}^\bot_i g(\tilde{\textbf{k'}}^\bot_i) |\Psi (\tilde{\textbf{k'}}^\bot_i)\rangle_s \langle \Psi (\tilde{\textbf{k'}}^\bot_i)|_s. \label{E:mixedstate}
\end{equation}

We refer to the angular spectrum of the signal photon, as conditioned by the detection of an idler photon, as the conditional angular spectrum (CAS).  It is given
by the following expression

\begin{equation}
\langle \hat{n}(\textbf{k}^\bot_s)\rangle=\mbox{Tr}( \hat{a}_s^\dagger(\textbf{k}^\bot_s)\hat{a}_s( \textbf{k}^\bot_s) \hat{\rho}_s)=\int d \tilde{\textbf{k'}}^\bot_i g(\tilde{\textbf{k'}}^\bot_i )|S(\textbf{k}^\bot_s+\tilde{\textbf{k'}}^\bot_i )|^2, \label{E:CAS}
\end{equation}

\noindent which is given by the displaced angular spectrum of the pump, averaged over the angular acceptance of the idler detector.  Note that in the idealized case discussed above of plane-wave idler detection, $g(\tilde{\textbf{k'}}^\bot_i )=\delta(\tilde{\textbf{k'}}^\bot_i-\tilde{\textbf{k}}^\bot_i)$ and the CAS becomes identical to the angular spectrum of the pump, except displaced

\begin{equation}
\langle \hat{n}(\textbf{k}^\bot_s)\rangle=|S(\textbf{k}^\bot_s+ \tilde{\textbf{k}}^\bot_i )|^2.\label{E:idealCAS}
\end{equation}

We are also interested in the transverse intensity of the heralded single photon, i.e. evaluated in the transverse position $\boldsymbol\rho^\bot_s$ rather than the transverse wavevector, defined in terms of annihilation operators in the transverse position domain $\tilde{a}_\mu(\boldsymbol\rho^\bot)=(4\pi^2)^{-1}\int d \textbf{k}^\bot e^{i \textbf{k}^\bot\cdot \boldsymbol\rho^\bot} \hat{a}_\mu( \textbf{k}^\bot)$, as

\begin{align}
\langle \tilde{n}(\boldsymbol\rho^\bot_s)\rangle=\mbox{Tr}( \tilde{a}_s^\dagger(\boldsymbol\rho^\bot_s) \tilde{a}_s( \boldsymbol\rho^\bot_s) \hat{\rho}_s)&=\int d \tilde{\textbf{k'}}^\bot_i g(\tilde{\textbf{k'}}^\bot_i )\left | 
\frac{1}{(2\pi)^2}
\int d \textbf{k}^\bot e^{i \textbf{k}^\bot\cdot \boldsymbol\rho^\bot_s}
S(\textbf{k}^\bot+\tilde{\textbf{k'}}^\bot_i )\right |^2. \label{E:transvintensity}
\end{align}

Applying the shift theorem, and noting that the phase $\mbox{exp}(i \rho^\bot_s\cdot \tilde{\textbf{k'}}^\bot_i )$ is suppressed by the absolute value, the integral over $\tilde{\textbf{k'}}^\bot_i$ becomes simply a constant $M$ and the single-photon transverse intensity reduces to 

\begin{equation}
\langle \tilde{n}(\boldsymbol\rho^\bot_s)\rangle=M \left | \tilde{S}(\rho^\bot_s) \right |^2,
\end{equation}

\noindent in terms of the Fourier transform of the pump transverse amplitude $\tilde{S}(\rho^\bot)$.  Thus, interestingly, the single-photon transverse intensity is identical in shape to the pump transverse intensity despite averaging over the angular acceptance of the idler detector.

Based on the physical mechanism discussed here, we have set out to experimentally demonstrate the transfer of a complex transverse structure from the pump to a heralded single photon.
In our experiment, see next section, we have used as pump in the SPDC process a BG beam, i.e a conical 
coherent superposition of Gaussian beams, each with a radius at the beam waist parameter $w_0$, and with a cone
opening half-angle  $\mbox{arctan}(k_t/k_p)$, where $k_t$ is the transverse wavenumber and $k_p$ is the pump wavenumber; the SPDC process with a Bessel beam pump has been studied  in  \cite{hernandez11,jeronimo14,prabhakar14}.  Note that in our experiment, in order to guarantee that the transverse amplitude of the pump is faithfully transferred to the heralded single photon, we
have selected a sufficiently small value of $k_t$ so that the short crystal approximation is valid (see discussion below).  For a BG beam of order $l$, the transverse amplitude can be written as

\begin{equation}\label{E:BGampl-k}
S(\textbf{k}^\bot)= A \exp\left(-\frac{w_0^2}{4} |\textbf{k}_\bot|^2\right) I_l\left(\frac{k_t w_0^2 |\textbf{k}_\bot|}{2}\right) \exp(i l \phi),
\end{equation}

\noindent in terms of the transverse wavevector $\textbf{k}^\bot\equiv(k_x,k_y)$, where $A$ is a normalization constant, $I_l(.)$ is an $l$th order
modified Bessel function of the first kind and $\phi=\mbox{arctan}(k_y/k_x)$~\cite{gutierrez05}.    The transverse amplitude for this BG beam
of order $l$ becomes 

\begin{equation}
\tilde{S}(\rho^\bot)= A^{'}\dfrac{1}{\mu} \exp \left\lbrace - \dfrac{1}{\mu} \left( \dfrac{ik^{2}_{t}z}{2k_{p}} + \dfrac{\mid \rho^\bot \mid ^{2}}{w^{2}_{0}} \right) \right\rbrace J_{l} \left( \dfrac{k_{t} \mid \rho^\bot \mid}{\mu} \right) 
\end{equation}

\noindent in terms of $\mu = 1+iz/z_{r}$ , with $z_{r} = k_{p} w^{2}_{0}/2$ the Rayleigh range of the pump and the propagation distance $z$; $J_{l}(.)$ is an $l$th order Bessel function of the first kind, and $A^{'}$  a normalization constant.
While in a previous paper~\cite{cruz12} we studied 
the single-photon non-diffractive behavior obtained from an $l=0$ BG pump, in this paper we concentrate on pump beams with OAM, specifically 
with orders $l=1$ and $l=2$.  

\begin{figure}[ht]
\centering
\includegraphics[width=12cm]{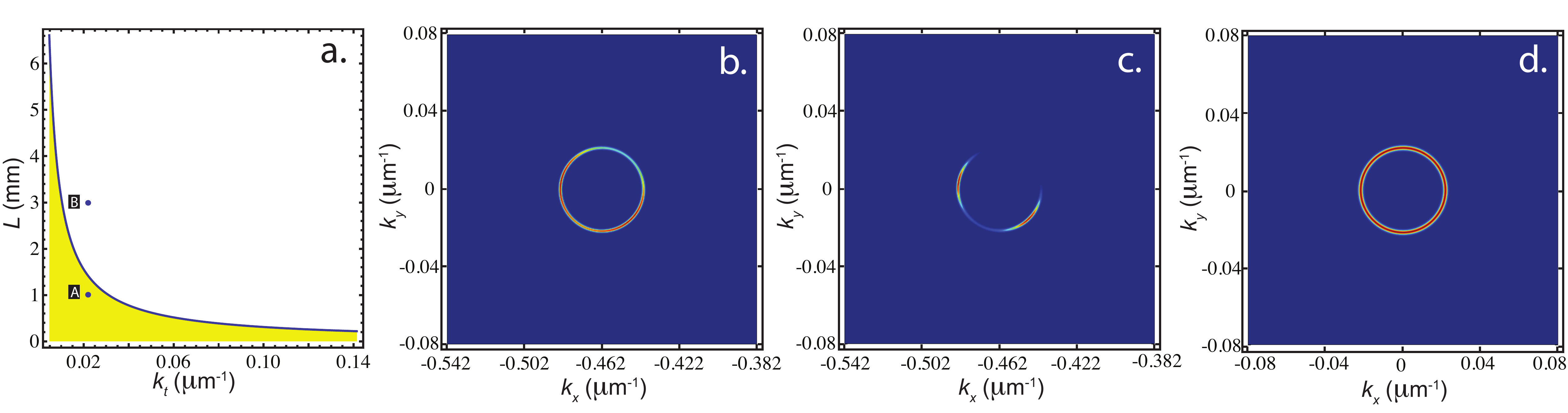}
\caption{(a) Parameter space (crystal $L$ and parameter $k_t$), with the region where the short crystal approximation is valid shaded in yellow.  While point A corresponds
to the experimental conditions (see below), point B lies outside of the short-crystal regime. (b) Simulated CAS for configuration A. (c) Simulated CAS for configuration B. (d) Assumed pump angular spectrum.
It may be seen that while the CAS for configuration A very nearly matches the pump angular spectrum, the CAS for configuration B differs substantially from the pump angular spectrum.}\label{Fig:shortcrystal}
\end{figure}

In the case of an SPDC source based on a BG pump there are two relevant parameters that determine whether or not the short crystal approximation is valid: 
the crystal length $L$ and the Bessel transverse wavenumber $k_t$.   Following the approach used in \cite{ramirez13},  in Fig.~\ref{Fig:shortcrystal}(a) we have shaded in yellow the region in the
$L$-$k_t$ parameter space where the short crystal approximation is valid.   The boundary of this region is determined as the set of values $\{L,k_t\}$ for which
the widths of the functions $|S( \textbf{k}^\bot_s+  \tilde{\textbf{k}}^\bot_i)|^2$ and $|G( \textbf{k}^\bot_s,  \tilde{\textbf{k}}^\bot_i,\omega_p-\tilde{\omega}_i )|^2$ become equal.  We
have indicated with a dot (labelled A) in this parameter space the parameters used in our experiment (see below; $L=1$mm, $k_t=2.2\times10^{-2}\mu$m$^{-1}$ and $w_0=1.3$mm) making it clear that the short crystal approximation is valid for our experimental
conditions.   In order to illustrate the effect of a departure from the short-crystal regime, we have also considered a configuration labelled B in Fig.~\ref{Fig:shortcrystal}(a) which corresponds to a longer crystal ($L=3$mm) with the remaining parameters unchanged.  
In Fig.~\ref{Fig:shortcrystal}(b) and (c) we have plotted the CAS as given by Eq.~\ref{E:idealCAS} for configurations A and B, respectively.  For comparison, in Fig.~\ref{Fig:shortcrystal}(d) we have plotted
the pump angular spectrum.   It is clear that while for configuration A
the CAS very nearly matches the pump angular spectrum, this is not the case for configuration B for which the function $|G( \textbf{k}^\bot_s,  \tilde{\textbf{k}}^\bot_i,\omega_p-\tilde{\omega}_i )|^2$
clips the function $|S( \textbf{k}^\bot_s+  \tilde{\textbf{k}}^\bot_i)|^2$.  It becomes clear that in this latter case, the pump angular spectrum is not faithfully transferred to the heralded single
photon.

In related work, it has been shown that the topological charge of a Laguerre-Gauss pump may be determined by the SPDC spatial coincidence count distribution~\cite{altman05}.   Note that our results differ from the appearance of a vortex in the joint spatial amplitude of a photon pair, as demonstrated in~\cite{gomes09}.  Note also that there is a separate class of experiments in which photon pairs may be prepared with a BG transverse structure, or even with entanglement in BG modes, with a pump which need not carry OAM, by appropriate post-generation projection of the signal/idler photons~\cite{mclaren12}.

In this paper we exploit a technique pioneered by Hickmann et al.~\cite{hickmann10,silva12}, through which the topological charge of an optical vortex beam can be revealed through the far-field diffraction pattern through a triangular aperture.  Specifically, it was found in the cited papers that the far-field diffraction pattern of an optical vortex with the phase singularity aligned with the center of a triangular aperture is formed by a triangular arrangement of intensity lobes, with the number of such lobes correlated to the topological charge.  The topological charge $m$ is then given by $N-1$ where $N$ is the number of lobes (discounting secondary lobes) on any side of the triangular intensity pattern.  In related work, a triangular slit arrangement has been used to verify the absence of three-silt interference and thus confirm the validity of Born's rule~\cite{hickmann11}.   As will be discussed below, we have used this technique to both: i)verify the presence of OAM, and ii) determine the topological charge, of heralded single photons obtained from the SPDC process implemented with a BG pump.

\section{Experiment}

\begin{figure}[ht]
\centering
\includegraphics[width=12cm]{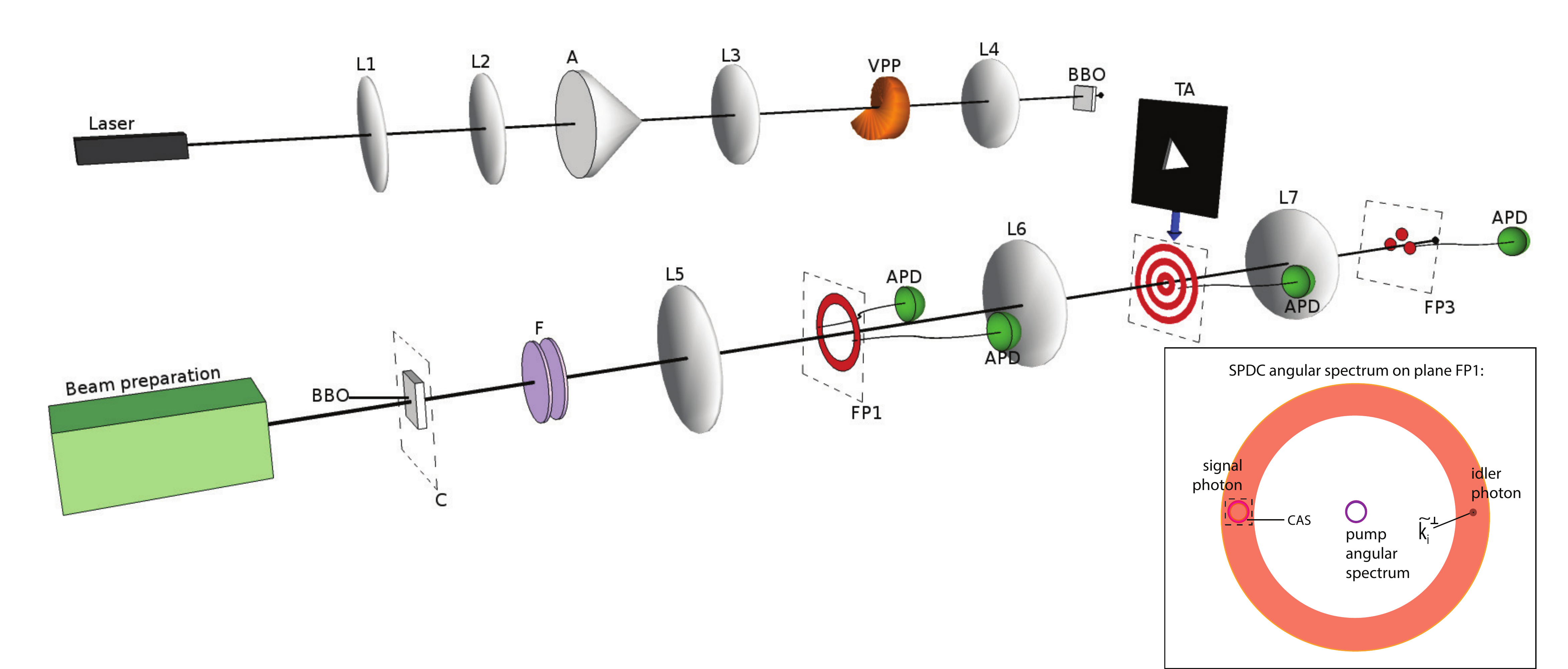}
\caption{Back: Bessel-Gauss beam preparation. Front:  Experimental setup for the transfer of optical vortices with $l=1$ and $l=2$ topological charge from a pump beam to a heralded single photon;
note that while all measurements use a spatially-resolving detector on FP1, a second detector is either not present (for the SPDC angular spectrum), or placed on any of the planes
FP1, FP2 and FP3 for measurements of the CAS, transverse intensity and far-field diffraction pattern, respectively.  Inset: SPDC and pump angular spectrum on plane FP1, where we have marked the location of the idler detector and the CAS; the dotted square represents the area where
the signal detector is scanned. }\label{Fig:setup}
\end{figure}

We use a BG beam, of order either 1 or 2, as pump in a second-order nonlinear crystal, which generates 
photon pairs through a type-I, non-collinear, frequency degenerate SPDC process. Figure~\ref{Fig:setup} shows our experimental setup.  
A beam from a diode laser, centered at $406$nm is sent through a telescope  with magnification $\times 10$, based on lenses L1 and L2 with focal lengths $5$cm and $50$cm,  to obtain
an approximately Gaussian beam with a radius at the beam waist of $\sim7.0$mm with power $\sim50$mW.   This magnified beam is transmitted
through an axicon (A), i.e. a conical lens, with a $1^\circ$ apex angle, which maps the incoming beam into a high-quality BG beam of 0th order.  
In order to turn this beam into a BG beam of order $1$ or $2$, we transmit the beam through a $\times 3$ telescope based on two lenses with focal
lengths $10$cm (L3) and $30$cm (L4) separated by $40$cm.   At the Fourier plane, i.e. at a distance of $10$cm from L3 we place 
a vortex phase plate (VPP), namely a phase mask with a linear azimuthal phase gradient covering either $0$ to $2 \pi$ phase for order $1$, or $0$ to 
$4 \pi$ for order $2$.  A BG beam with the desired optical vortex structure thus forms after L4.

\begin{figure}[ht]
\centering
\includegraphics[width=12cm]{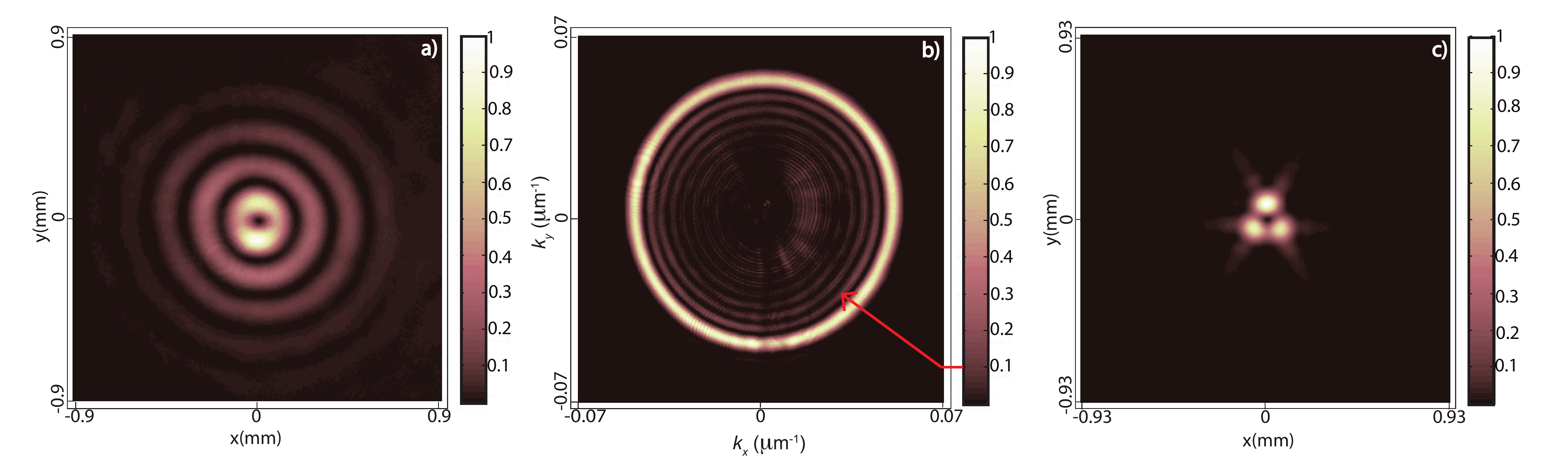}
\caption{For a Bessel-Gauss optical vortex with topological charge $l=1$: (a) Transverse intensity pattern, (b) angular spectrum, (c) far-field diffraction pattern through an triangular aperture}\label{Fig:bombeo1}
\end{figure}

\begin{figure}[ht]
\centering
\includegraphics[width=12cm]{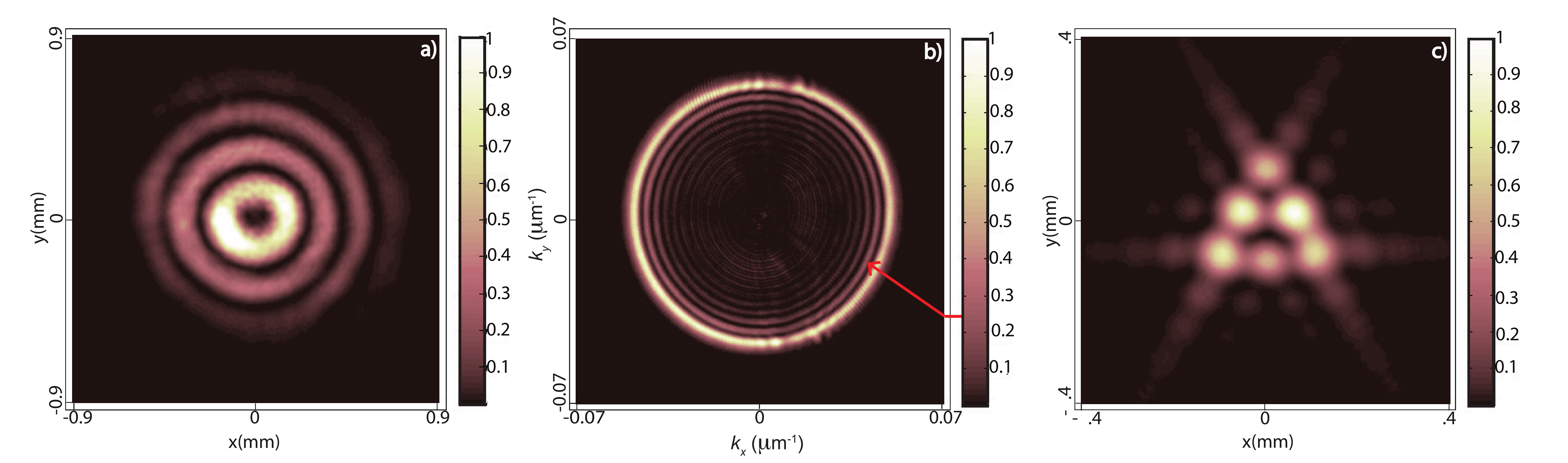}
\caption{For a Bessel-Gauss optical vortex with topological charge $l=2$: (a) Transverse intensity pattern, (b) angular spectrum, (c) far-field diffraction pattern through a triangular aperture.}\label{Fig:bombeo2}
\end{figure}

Figure~\ref{Fig:bombeo1}(a) shows, for the BG pump beam of order $1$, a measurement of the transverse intensity, obtained with a CCD camera, at a distance of $16$cm from L4 (see Fig.~\ref{Fig:setup}); this will be referred to as the C plane.    This measurement
shows the distinct characteristics of a higher order ($l \ge 1$) BG vortex beam, i.e. a central null in intensity surrounded by Bessel rings.  The slight departure from azimuthal symmetry is probably
due to fabrication imperfections in the axicon and/or the VPP.  Figure~\ref{Fig:bombeo1}(b) shows
the corresponding transverse momentum distribution, obtained through an optical $f$-$f$ system prior to the CCD camera.    This figure shows the expected
annular shape of the BG pump angular spectrum;   the structure observable within the ring (with maximum intensity around $10\%$ of that of the main ring, as indicated in red) can be attributed, again, to imperfections in the axicon and/or VPP.   In order to verify the vortex structure of this beam, we have measured the far-field diffraction pattern obtained from an equilateral triangular aperture (with sides of $500\mu$m length).    A CCD camera on the far field plane, corresponding to the Fourier plane following an $f$-$f$ optical system, yields a three-lobe intensity pattern (with some side-lobes) [see Fig.~\ref{Fig:bombeo1}(c)], characteristic of an optical vortex with topological charge $1$.  Figure~\ref{Fig:bombeo2} is similar to Fig.~\ref{Fig:bombeo1}, for a BG pump beam of order $2$; the diffraction pattern through a triangular aperture in this case yields a six-lobe intensity pattern (again with some side-lobes), characteristic of an optical vortex with topological charge $2$.  Note that the radius of the annulus in Figs.~\ref{Fig:bombeo1}(b) and \ref{Fig:bombeo2}(b),
in $\textbf{k}^\bot$ space, directly yields the parameter $k_t$ of Eq.~(\ref{E:BGampl-k}), in our case with a value of  $k_t=2.2 \times 10^{-2} \pm .001 \mu \mbox{m}^{-1}$.  Note also that the
$w_0$ parameter in Eq.~(\ref{E:BGampl-k}) may be obtained from the width of the annulus $\delta k$ according to the relationship $w_0=4/\delta k$; in our case we obtain a value of
$w_0=1.3 \pm 0.3 $mm.

\begin{figure}[ht]
\centering
\includegraphics[width=10cm]{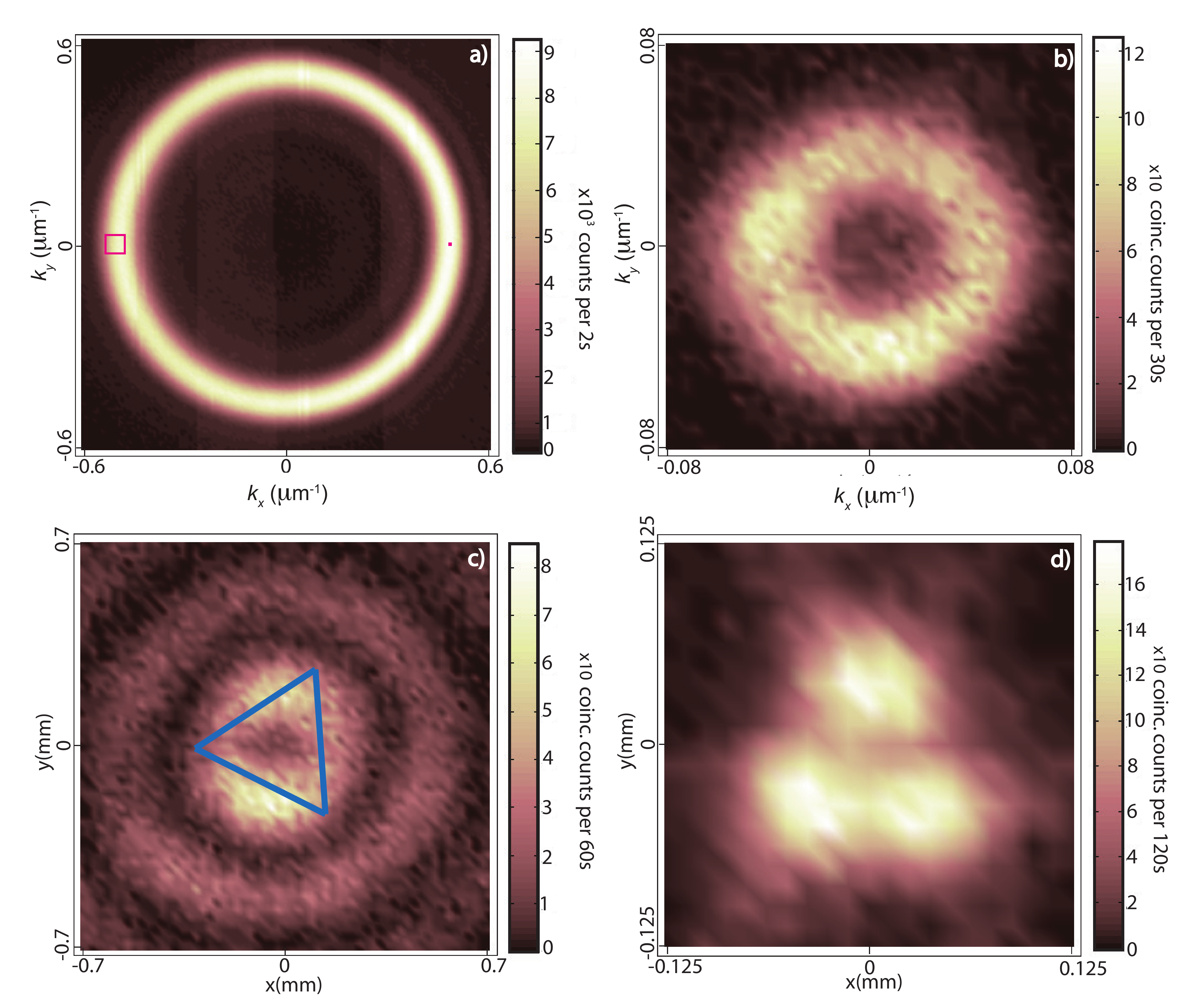}
\caption{For a pump beam with topological charge $l=1$: (a) SPDC angular spectrum, (b) conditional angular spectrum of the signal-mode single photons, (c) transverse
intensity distribution of signal-mode heralded single photon, (d) far-field diffraction pattern through a triangular aperture (sketched in panel c) of signal-mode heralded single photon. }\label{Fig:bessel1}
\end{figure}

The SPDC crystal, $\beta$-barium borate (BBO) with thickness  $1$mm, is placed on the C plane (see Fig.~\ref{Fig:setup}) .  Pump photons are eliminated using appropriate filters (F): a $\lambda>490$nm long wave pass filter followed by a bandpass filter centered at $810$nm with a $40$nm bandwidth.
An  $f$-$f$ optical system is used in order to yield a Fourier plane on which we can probe the signal and idler
transverse momentum distributions.   Specifically, a lens (L5; focal length $f=10$cm and $1$-inch diameter)  which is placed at a distance of $10$cm 
from the C plane, defines a Fourier plane (FP1) at a distance of $10$cm from the lens.  The typical cone of type-I SPDC non-collinear photon pairs, resolved in transverse momentum space on this Fourier plane, is then referred to as the SPDC angular spectrum.  Spatially-resolved photon counting on FP1 thus yields the angular spectrum of the SPDC photon pairs, convolved with the angular acceptance function of the detector.   For this purpose, we have used a fiber tip of a large-diameter fiber ($200\mu$m) which can be 
displaced laterally along the $x$ and $y$ directions with the help of a computer-controlled motor ($50$nm resolution and $1.5$cm travel range).  The fiber used leads
to a Si avalanche photodiode (APD), with its output connected  to standard pulse-counting equipment to obtain the number of detection events per unit time; Fig.~\ref{Fig:bessel1}(a)
shows the resulting measured SPDC angular spectrum.    

Because the signal and idler photons of a given pair reach FP1 on opposite sides of the SPDC angular spectrum, they may be detected in coincidence through separate spatially-resolved detectors.  By leaving
the idler detector fixed while scanning the signal detector, we obtain a measurement the signal photon CAS   $\langle \hat{n}(\textbf{k}^\bot_s)\rangle$  (corresponding to Eq.~\ref{E:CAS}),
convolved with the angular acceptance function of the scanning detector.   The inset in Fig.~\ref{Fig:setup} shows the pump and the SPDC angular spectra, as well as the location of the idler detector on the right hand side of the ring and
the signal photon with a CAS which matches in shape the pump angular spectrum on the left hand side of the ring.  Figure~\ref{Fig:bessel1}(b) shows this measured CAS, i.e.  the angular spectrum of the signal photon conditioned on the detection of an idler photon with a transverse momentum value defined by the position of the idler detector.
Specifically, the idler fiber tip is placed at the disk indicated in Fig.~\ref{Fig:bessel1}(a) (see right hand side of angular spectrum), corresponding to $k^\bot_{iy}=0$ and with a $k^\bot_{ix}$ value which maximizes the counts, while  we have scanned the signal-mode fiber tip, also with a $200\mu$m diameter,  around the diametrically-opposed point, within the square indicated in Fig.~\ref{Fig:bessel1}(a).  Note that the fiber used for idler-photon collection is highly multi-mode 
and its angular acceptance function corresponds to the function $ g(\tilde{\textbf{k'}}^\bot_i) $ in Eq.~(\ref{E:mixedstate}).  Detection over many transverse fiber modes leads to the incoherent sum over detection idler transverse wavectors in Eq.~(\ref{E:mixedstate}).    The plot in  Fig.~\ref{Fig:bessel1}(b) corresponds to the number of signal-idler measured coincidence counts as a function of the position (transverse momentum) of the signal-mode fiber tip.      This plot shows, as expected from Eq.~(\ref{E:transfer}), an annular structure which for an idealized vanishing angular idler-photon collection width would be identical to the pump angular spectrum, plotted in Fig.~\ref{Fig:bombeo1}(b).  The larger single-photon conditional angular spectrum width, compared to the pump angular spectrum width, is related to the significant width of the conditioning and scanning fiber tips.

\begin{figure}[ht]
\centering
\includegraphics[width=10cm]{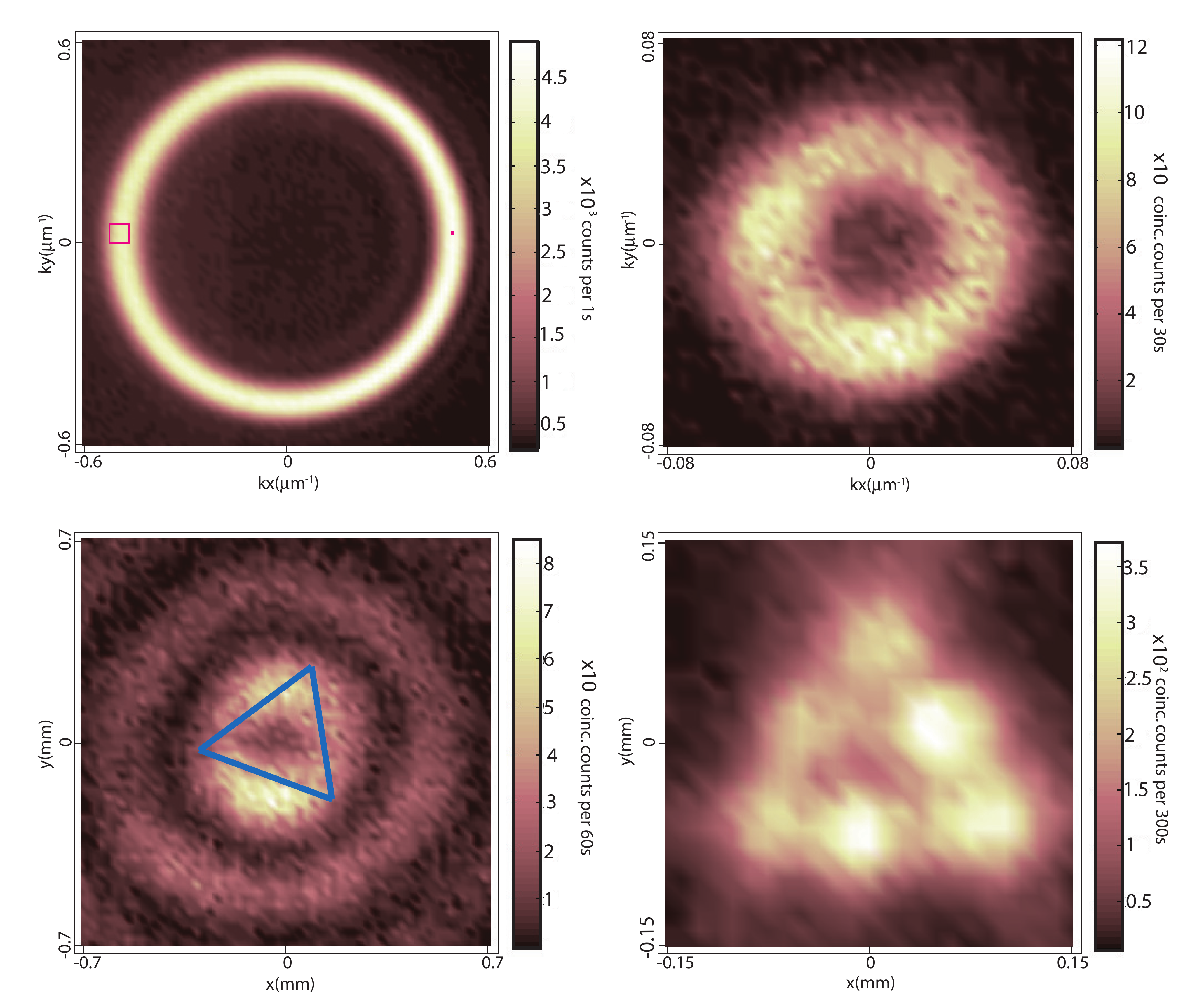}
\caption{For a pump beam with topological charge $l=2$: (a) SPDC angular spectrum, (b) conditional angular spectrum of the signal-mode single photons, (c) transverse
intensity distribution of signal-mode heralded single photon, (d) far-field diffraction pattern through a triangular aperture (sketched in panel c) of signal-mode heralded single photon. }\label{Fig:bessel2}
\end{figure}

Besides the conditional angular spectrum of the signal photon, shown in Fig.~\ref{Fig:bessel1}(b), we are interested in measuring the transverse single-photon intensity  $\langle n(\boldsymbol\rho^\bot_s)\rangle$ (see Eq.~\ref{E:transvintensity}).   For this purpose, we have placed a second $f$-$f$ optical system following FP1.  Specifically, we have placed a lens (L6; focal length $f_2=15$cm and  2-inch diameter) at a distance of $15$cm from FP1, so as to define a second Fourier plane (FP2) a distance of $15$cm from the lens (see Fig.~\ref{Fig:setup});   the larger lens diameter lens is intended to minimize signal-mode $k$-vector clipping by the lens aperture.   For this measurement,  we retain the fixed conditioning idler detector on FP1 and we place the signal-mode fiber tip with a $50\mu$m diameter, rather than $200\mu$m as for FP1,  and scan its transverse position while monitoring signal-idler coincidence counts.     The resulting data, i.e coincidence counts between idler photons collected on FP1 and signal photons collected on FP2 as a function of the position of the signal-mode fiber tip constitutes a measurement of the heralded  signal-mode intensity (convolved with the angular acceptance of the scanning detector) as a function of transverse \emph{position} rather than \emph{momentum}.

In addition, we are interested in verifying the vortex nature of the heralded signal photon.  For this purpose, we place an equilateral triangular aperture (TA) with sides of $500\mu$m length on FP2.  Note that the position, dimensions and orientation of TA are indicated in blue in Fig.~\ref{Fig:bessel1}(c).  The signal-mode single photon diffracts through this aperture; in order to be able to probe the diffracted far-field transverse intensity pattern, we place a lens (L7; focal length $f=3$cm) a distance of $3$cm from  FP2, so as to define a third Fourier plane, FP3, a distance of $3$cm from the lens.     For this measurement,  we retain the fixed conditioning idler detector on FP1 and we place the signal-mode fiber tip with a $50\mu$m diameter,  and scan its transverse position while monitoring signal-idler coincidence counts.     The resulting data, i.e coincidence counts between idler photons collected on FP1 and signal photons collected on FP3 as a function of the position of the signal-mode fiber tip constitutes a measurement of the far-field diffracted heralded signal-mode transverse intensity.   The resulting three-lobe pattern, very similar to that obtained for the pump (see Fig.~\ref{Fig:bombeo1}(c)), constitutes a clear indication that the signal-mode heralded single photon is a vortex wave.  Thus, we have demonstrated the transfer of the classical vortex pump wave to the transverse amplitude profile of the heralded single photon.    

We have carried out the above experiment, both, for 1st and 2nd order BG beams.   Figure~\ref{Fig:bessel2} is similar to Fig.~\ref{Fig:bessel1}, for a 2nd order BG beam with properties shown in Fig.~\ref{Fig:bombeo2}.  Once again, for this 2nd order BG experiment, diffraction through a triangular aperture demonstrates  the expected transfer vortex in the classical  pump to the heralded single photon. 

Note that the experiment may be interpreted in terms of Klyshko's advanced wave picture~\cite{klyshko88} in which the crystal behaves as a mirror and the conditioning detector as a source; this brings some parallels with a ghost diffraction setup~\cite{strekalov95,kang10,tasca13}.   If we consider the fixed idler detector on plane FP1 now a point source, propagation backwards towards the crystal through lens L5 collimates the rays which 
reach the plane of the crystal and are modulated in amplitude and in phase by a mask (which corresponds to the transverse amplitude of the pump beam) on the plane of the crystal.   The rays now propagate through a succession of two lenes, the first of which implements a Fourier transform and the second an inverse Fourier transform, thus yielding the same transverse amplitude on plane FP2 as on the plane of the mask (crystal); the intensity is either directly probed on FP2 by spatially resolved detection or studied through the far field diffraction pattern through a triangular aperture.

\section{Conclusions}

We have reported an experiment in which an SPDC crystal is pumped by a higher-order Bessel Gauss beam with orbital angular momentum, specifically with topological charge $l=1$ and $l=2$. We have experimentally verified that, in the the short-crystal regime in which our source operates, the transverse properties of the pump are transferred to each signal-mode single photon, as heralded by the detection of the corresponding idler photon.    We have verified this transfer by measuring, for the heralded single photon: i)the conditional angular spectrum, ii) the transverse intensity, and iii) the far-field diffraction pattern through a triangular aperture.  In all three cases, we obtained results which match well the corresponding measurements carried out for the pump.  In particular, the three-lobe single-photon diffraction pattern obtained for an $l=1$ pump and the six-lobe diffraction pattern obtained for an $l=2$ pump serve as confirmation that our heralded single photon carries OAM with topological charge $1$ and $2$ respectively.

\section*{Acknowledgments}

This work was supported by CONACYT, Mexico, by DGAPA (UNAM) and by AFOSR grant FA9550-13-1-0071.

\end{document}